\documentclass[12pt]{article}
\usepackage[dvips]{graphics}
\usepackage{epsfig}

 \newcommand\la{\langle}
 \newcommand\ra{\rangle}
 \newcommand\beq{\begin{equation}}
 
 \newcommand\eeq{\end{equation}}
 \newcommand\beqn{\begin{eqnarray}}
 \newcommand\eeqn{\end{eqnarray}}

\def\mb{\,\mbox{mb}}
\def\fm{\,\mbox{fm}}
\def\GeV{\,\mbox{GeV}}
\def\TeV{\,\mbox{TeV}}
\def\Pom{{\bf I\!P}}

\def\lsim{\mathrel{\rlap{\lower4pt\hbox{\hskip1pt$\sim$}}
    \raise1pt\hbox{$<$}}}
\def\gsim{\mathrel{\rlap{\lower4pt\hbox{\hskip1pt$\sim$}}
    \raise1pt\hbox{$>$}}}

\begin{document}

\date{}

\title{\bf Unusual Features of\\ Drell-Yan Diffraction}

\maketitle

\begin{center}

\vspace*{-1.5cm}

\author{B.Z.~Kopeliovich$^{1-3}$, I.K.~Potashnikova$^{1,3}$,
Ivan~Schmidt$^{1}$ and A.V.~Tarasov$^{2,3}$}

\vspace*{0.5cm}

$^{1}$ {Departamento de F\'\i sica, Universidad T\'ecnica Federico Santa
Mar\'\i a, Casilla 110-V, Valpara\'\i so, Chile}

$^2$ {Institut f\"ur Theoretische Physik der Universit\"at,
Philosophenweg 19, 69120 Heidelberg, Germany}

$^{3}$Joint Institute for Nuclear Research, Dubna,
Russia

\end{center}

\begin{abstract}
 The cross section of the diffractive Drell-Yan (DY) process, $pp\to
\bar ll Xp$, where the system $\bar ll X$ is separated by a large
rapidity gap from the recoil proton, is calculated in the light-cone
dipole approach. This process reveals unusual features, quite
different from what is known for diffractive DIS and nonabelian
radiation: (i) the diffractive radiation of a heavy dilepton by a
quark vanishes in the forward direction; (ii) the diffractive
production of a dilepton is controlled by the large hadronic radius;
(iii) in contrast with DIS where diffraction is predominantly soft,
the diffractive DY reaction is semihard-semisoft; (iv) as a result of
the saturated shape of the dipole cross section, the fraction of
diffractive DY events steeply falls with energy, but rises as function
of the hard scale. These features are common for other abelian
bremsstrahlung processes (higgsstrahlung, Z-strahlung, etc.).
Measurements of diffractive DY processes at modern colliders 
would be a sensitive probe for the shape of the dipole cross 
section at large separations.

 \end{abstract}



\section{Diffractive radiation: heuristic approach}

Diffraction excitation of hadrons is possible due to the presence
of quantum fluctuations in the projectile particles. In classical
physics only elastic diffraction is possible. It was first
realized by Feinberg and Pomeranchuk \cite{fp} and Good and Walker
\cite{gw} that the compositeness of hadrons leads to production of
new states. Although different Fock components of the hadron
experience only elastic scattering, which is a shadow of inelastic
collisions, the wave packet composition may be altered producing a
new hadronic state. Indeed, this may happen if the Fock states
interact differently, otherwise the wave packet retains the same
composition, i.e. the final and initial states are identical.

The dipole description of diffraction in QCD was presented in
\cite{zkl,bbgg}. Since dipoles of different transverse size $r_T$
interact with different cross sections $\sigma(r_T)$, this gives
rise to single inelastic diffraction with a cross section given by
the dispersion of the $r_T$-distribution \cite{zkl},
 \beq
\left.\frac{d\sigma_{sd}}{dp_T^2}\right|_{p_T=0} =
\frac{\la\sigma^2(r_T)\ra -
\la\sigma(r_T)\ra^2}{16\,\pi}\ .
\label{10}
 \eeq
 Here $p_T$ is the transverse momentum of the recoil proton;
$\la\sigma(r_T)\ra$ is the dipole-proton cross section averaged over
dipole separation.

The dipole description of diffractive radiation of photons and
gluons was developed in Ref. \cite{kst2}. Diffractive radiation of
a photon by a quark, in which the photon can be either real, or
heavy decaying into a dilepton, turns out to vanish in the forward
direction. Indeed, let us consider two Fock components of a quark,
just a bare quark, $|q\ra$, and a quark accompanied by a
Weizs\"acker-Williams photon, $|q\gamma^*\ra$. In both components
only the quark can interact, therefore the two terms in (\ref{10})
cancel each other. Notice that the partial diffractive amplitude
at a given impact parameter does not vanish, since the recoil
quark in the $|q\gamma^*\ra$ state gets a shift in impact
parameters compared to the $|q\ra$ state, and the two Fock
components interact differently. Only after integration over
impact parameter, corresponding to the forward amplitude, the
diffractive radiation vanishes.

\begin{figure}[htbp]
 \includegraphics[width=12cm]{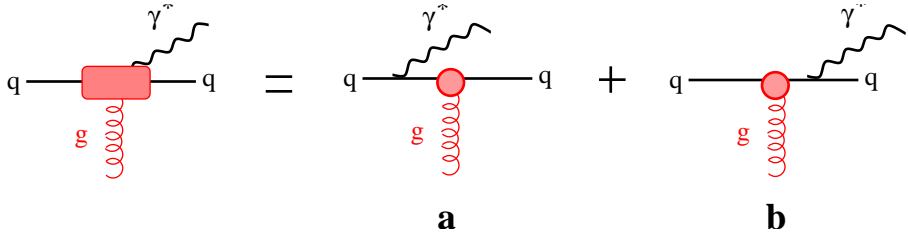}
 \caption{Born approximation for radiation of a photon by a quark in
inelastic collision}
 \label{graphs1}
 \end{figure}

A direct calculation of Feynman graphs \cite{kst2} confirms this
expectation. Notice that even nondiffractive photon radiation in
inelastic collisions is impossible without momentum transfer.
Indeed, the Born graphs Fig.~\ref{graphs1}a and \ref{graphs1}b
cancel if $p_T=0$.
 Intuitively it is clear that if the electric charge gets no kick and is
not accelerated, then no radiation happens.

This result is retained in the nonabelian case, namely a quark does
not radiate gluons if the t-channel gluon provides no momentum
transfer \cite{gb}. This is not a trivial result, since in this
case a color current flows between the beam and target.

\begin{figure}[htbp]
 \includegraphics[width=13.5cm]{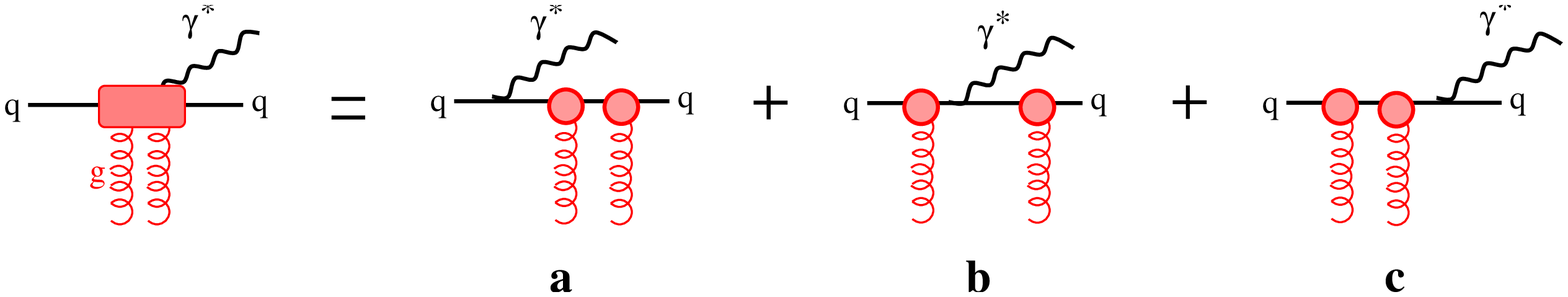}
 \caption{Born graphs for diffractive electromagnetic radiation by a
quark.}
 \label{graphs2}
 \end{figure}

In the case of {\it diffraction} forward electromagnetic radiation
vanishes as well, but this is less obvious. Indeed, even if the two-gluon
exchange provides no momentum transfer, each of gluons can carry a
transverse momentum. The relevant Born graphs for diffractive photon
radiation are shown in Fig.~\ref{graphs2}.
 Graph (b) does not contribute to radiation, provided that the radiation
time considerably exceeds the duration time of interaction.  Only
graphs (a) and (c) can contribute, but they cause no radiation for
forward scattering for the same reason as in the inelastic
collision, Fig.~\ref{graphs1}.

This can be interpreted as a consequence of Landau-Pomeranchuk principle
\cite{lp}: radiation depends on the whole strength of the kick, rather
than on its structure, if the time scale of the kick is shorter than the
radiation time. In other words, if two opposite kicks are separated by a
short time interval, the radiation spectra from each of the two kicks
interfere destructively and compensate each other \cite{kz,feri}, i.e. no
radiation occurs.

Notice that disappearance of abelian diffractive radiation in the forward
direction goes along with the result of \cite{bln} that the diffractive
cross section is proportional to the mean momentum transfer squared. That
was, however, an oversimplified treatment of the Pomeron as a point-like
vector meson.

Although a quark cannot radiate a dilepton in forward direction, a hadron
can. That is possible due to transverse motion of the valence quarks in the
hadron, i.e. abelian radiation even at a hard scale is sensitive to the
hadron size, which is a dramatic break down of QCD factorization
\cite{landshoff} (which has never been proven for this process).  Failure
of factorization for diffractive Drell-Yan reaction has been already known.
It was found in \cite{cfs,collins} that factorization fails due to the
presence in the Pomeron of spectator partons. Below we demonstrate that
factorization in Drell-Yan diffraction is even more broken due to presence
of spectator partons in the colliding hadrons.

Notice that diffractive {\it nonabelian} (gluon) radiation is different,
it does not vanish in the forward direction \cite{kst2}. This is a direct
manifestation of the nonabelian dynamics. In addition to the three graphs
for electromagnetic radiation shown in Fig.~\ref{graphs2}, in the case of
gluon radiation there is a fourth term corresponding to one of the
t-channel gluons coupled to the radiated one. This term gives rise to a
nonzero forward diffraction. This is confirmed by data, since a nonzero
forward cross section of diffractive gluon radiation, called
triple-Pomeron term in Regge approach, is well established experimentally
\cite{3R}. Nevertheless, the cross section of diffractive gluon radiation
turns out to be amazingly small. Indeed the Pomeron-proton total cross
section extracted from data for large mass diffraction turns out to be
only $2\mb$, an order of magnitude smaller than for pion-proton. This
effect is discussed and explained in Ref. \cite{kst2}.

\section{Calculation of Feynman graphs}

Although a quark cannot diffractively radiate photons in forward
direction, a hadron can.  Let us consider a $pp$ collision,
 \beq
p_b+p_t \to \gamma^* + X + p_t\ ,
\label{20}
 \eeq
 assuming that the initial beam nucleon and the final state consist of
three valence quarks, $p_b={3q}_{i}$; $X={3q}_{f}$. The amplitude of
the Drell-Yan (DY) reaction Eq.~(\ref{20}) has three terms,
 \beq
 A_{if}= A_{if}^{(1)}+ A_{if}^{(2)}+ A_{if}^{(3)}\ ,
\label{30}
 \eeq
 where subscripts $i$ and $f$ correspond to initial and final states of 
the Drell-Yan diffractive reaction, Eq.~(\ref{20}).

\begin{figure}[htbp]
 \includegraphics[width=13.5cm]{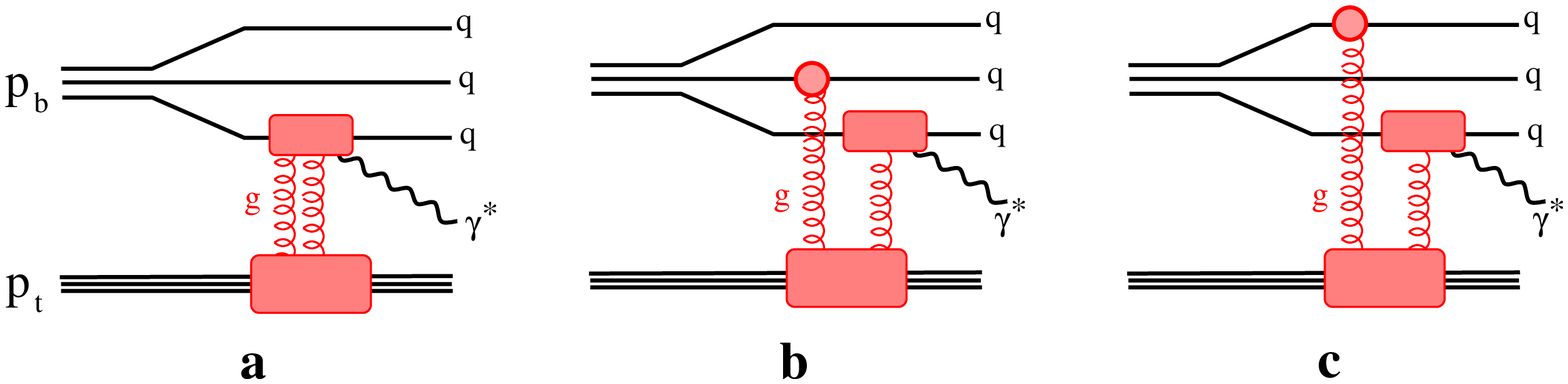}
 \caption{Born graphs for diffractive electromagnetic radiation by a
proton}
 \label{graphs3}
 \end{figure}

 Each term in (\ref{30}) corresponds to radiation of the $\gamma^*$ by one
of the valence quarks, and it includes contributions from tree graphs
shown for $A_{if}^{(1)}$ in Fig.~\ref{graphs3}.
 The first graph (a) leads to no radiation at $p_T=0$ as was
explained above. The rest, graphs (b) and (c), can be calculated according
to,
 \beqn
&& A_{if}^{(1)}(x_\gamma,\vec k)\Bigr|_{p_T= 0} 
= \frac{i}{8\pi}\int d^2r_1
d^2r_2 d^2r_3 d^2r dx_{q_1} dx_{q_2} dx_{q_3}
 \nonumber\\ &\times&
\Psi_i(\vec r_1,\vec r_2,\vec r_3;x_{q_1},x_{q_2},x_{q_3})
\Psi^*_f(\vec r_1+\alpha\vec r,\vec r_2,\vec
r_3;x_{q_1}-x_\gamma,x_{q_2},x_{q_3}) \nonumber\\ &\times&
\Sigma^{(1)}(\vec r_1,\vec r_2,\vec r_3,\vec r,\alpha)\,
\Phi_1(\vec r,\alpha)\, e^{-i\vec k\cdot\vec r}\ .
 \label{40}
 \eeqn
 Here $\Psi_{i,f}$ are the light-cone wave functions of the $3q$ systems
in the initial and final state respectively; $\vec r_1,\ \vec r_2$
and $\vec r_3$ are the impact parameters of the quarks;
$x_{q_1},x_{q_2},x_{q_3}$ are the fractions of the proton
light-cone momentum carried by the quarks; $\vec k$ and $x_\gamma$
are the photon transverse and fractional longitudinal momenta;
$\vec r$ is the transverse separation between the photon and the
radiating quark; $\alpha=x_\gamma/x_{q_1}$; and $\Phi_i(\vec
r,\alpha)$ is the distribution amplitude for photon radiation by
quark $q_i$ in the mixed representation, transverse coordinate and
longitudinal momentum. The factor $\Sigma^{(1)}$ has the form,
 \beq
\Sigma^{(1)}(\vec r_1,\vec r_2,\vec r_3,\vec r,\alpha) =
\sigma(\vec r_1-\vec r_2)-
\sigma(\vec r_1-\vec r_2-\alpha\vec r)+
\sigma(\vec r_1-\vec r_3)-
\sigma(\vec r_1-\vec r_3-\alpha\vec r)\ ,
\label{50}
 \eeq
 where $\sigma(r_T)$ is the universal dipole cross section \cite{zkl}.

The structure of this amplitude is easy to understand. According to the
discussion in the previous section, the amplitude of diffractive radiation
should be proportional to the difference between elastic amplitudes of the
two Fock components, with and without the photon. In both cases only the
3-quark dipoles interact, but they have different sizes in the $|3q\ra$
and $|3q\gamma^*\ra$ components, therefore interact differently.  Indeed,
when a quark fluctuates into quark-photon with transverse separation $\vec
r$, the final quark gets a transverse shift $\Delta \vec r = \alpha\vec
r$, where $\alpha$ is the fraction of the quark momentum taken away by the
photon. This shift is explicitly presented in (\ref{50}).

For the model of symmetric proton wave function we use, the interference
terms in the cross section cancel, since they are proportional to
$Z_1Z_2+Z_1Z_3+Z_2Z_3=0$ (no cancellation occurs for neutrons), where
$Z_i$ is the electric charge of the $i$th quark.  Thus, summing over
different final states $f$ one gets for the forward diffractive cross 
section,
 \beqn
\left.\frac{d\sigma_{sd}^{DY}}{dp_T^2}\right|_{p_T=0} &=&
\frac{1}{64\pi}\sum\limits_f\int d^2k\,
\frac{dx_\gamma}{x_\gamma}
\sum\limits_{n=1}^3
\left|A^{(n)}_{if}\right|^2
\nonumber\\ &=&
\frac{1}{64\pi}\sum\limits_f\int d^2k\,
\frac{dx_\gamma}{x_\gamma}
\,\left|A^{(1)}_{if}\right|^2
\left(1+\frac{Z_2^2}{Z_1^2} +
\frac{Z_3^2}{Z_1^2}\right)\ .
\label{60}
 \eeqn

Employing completeness,
 \beqn
&& \sum\limits_f
\Psi_f(\vec r_1,\vec r_2,\vec r_3;x_{q_1},x_{q_2},x_{q_3})
\Psi^*_f(\vec r_1^{\,\prime},\vec r_2^{\,\prime},\vec r_3^{\,\prime};
x_{q_1}^\prime,x_{q_2}^\prime,x_{q_3}^\prime)
\nonumber\\ &=&
\prod\limits_{j=1}^3\delta(\vec r_j-\vec r_j^{\,\prime})
\delta(x_{q_j}-x_{q_j}^\prime)\,
\label{70}
 \eeqn
 we arrive at
 \beqn
\left.\frac{d\sigma_{sd}^{DY}}{dp_T^2}\right|_{p_T=0} &=&
\frac{\sum Z_q^2}{64\pi}
\int d^2r_1 d^2r_2 d^2r_3 d^2r\,
dx_{q_1} dx_{q_2} dx_{q_3}\,
\frac{dx_\gamma}{x_\gamma}\,
\left|\tilde\Phi(\vec r,\alpha)\right|^2
\nonumber\\ &\times&
\left|\Psi_i(\vec r_1,\vec r_2,\vec r_3;x_{q_1},x_{q_2},x_{q_3})\right|^2
\left|\Sigma^{(1)}(\vec r_1,\vec r_2,\vec r_3,\vec r,\alpha)\right|^2\ .
\label{80}
 \eeqn
 Here $\tilde\Phi(\vec r,\alpha)=\Phi_i(\vec r,\alpha)/Z_i$ is the
amplitude of photon radiation by a quark with charge one.

To progress further we assume the Gaussian shape for the quark
distribution in the proton,
 \beqn
\left|\Psi_i(\vec r_1,\vec r_2,\vec r_3;x_{q_1},x_{q_2},x_{q_3})\right|^2
&=& \frac{3a^2}{\pi^2}\,\exp[-a(r_1^2+r_2^2+r_3^2)]
\rho(x_{q_1},x_{q_2},x_{q_3})
\nonumber\\ &\times&
\delta(\vec r_1+\vec r_2+\vec r_3)\,
\delta(1-x_{q_1}-x_{q_2}-x_{q_3})\ ,
\label{90}
 \eeqn
 where $a=\la r_{ch}^2\ra^{-1}$ is the inverse proton mean charge radius
squared. The three-body distribution function
$\rho(x_{q_1},x_{q_2},x_{q_3})$ should
reproduce the valence quark distribution in the proton,
 \beq
\int dx_{q_2}dx_{q_3}\,\rho(x_{q_1},x_{q_2},x_{q_3}) =
\rho_{q_1}(x_{q_1})\ .
\label{100}
 \eeq

Summing different quark and antiquark species one arrives at the proton
structure function \cite{krt3},
 \beq
\sum\limits_{q} Z_q^2\,
\left[\rho_{q}(x)+\rho_{\bar q}(x)\right]
= {1\over x}\,F_2(x)\ .
\label{105}
 \eeq
 At this point we generalize our three-body quark wave function (\ref{90})
to a multi-body one including antiquarks. However, we keep the same simple
coordinate part of the wave function.

In what follows we rely on the popular saturated form of the dipole cross
section,
 \beq
\sigma(r)=\sigma_0\left(1-e^{-r^2/R_0^2}\right)\,
\label{110}
 \eeq
 where the parameters fitted to DIS data at small $x$ can be found
in \cite{gbw}.

The distribution functions have the form,
 \beq
\tilde\Phi^{L}(\vec r,\alpha)=
\frac{\sqrt{\alpha_{em}}}{\pi}\,M(1-\alpha)\,
 \chi^\dagger_f\,\chi_i\,
K_0(\tau r)\ ;
\label{120}
 \eeq
 \beqn
\tilde\Phi^{T}(\vec r,\alpha) &=&
\frac{\sqrt{\alpha_{em}}}{2\pi}\,
\chi^\dagger_f\,\left\{
i\alpha^2 m_q [\vec\sigma\times\vec e]\cdot\vec n
-i(2-\alpha)\, \vec e\cdot\vec\nabla
\right.\nonumber\\ &-& \left.
\alpha [\vec\sigma\times\vec e]\cdot\vec\nabla\right\}\chi_i\,
K_0(\tau r)\ ,
\label{125}
 \eeqn
 for radiation of longitudinally and transversely polarized photons
respectively. Here $K_0(\tau r)$ is the modified Bessel function;
 \beq
\tau^2=M^2(1-\alpha) + m_q^2\,\alpha^2\ ;
\label{130}
 \eeq
 $\chi_{i,f}$ are the spinors assigned to the quark before and after the
radiation respectively; $\vec\sigma$ is the Pauli matrix; $\vec n$ is a
unit vector directed along the initial collision momentum; $m_q$ is the
quark mass.

Now we are in a position to perform integration over coordinates in
(\ref{80}). Using the integral representation for the modified Bessel
function,
 \beqn
K_0(\tau r) &=& {1\over2}\int\limits_0^\infty
\frac{dt}{t}\,
\exp\left(-\frac{\tau^2 r^2}{4t}\right)\ ;
\nonumber\\
K_0^2(\tau r) &=& {1\over4}\int\limits_0^\infty
\frac{dt}{t}\int\limits_0^\infty \frac{du}{u}\,
\exp\left[-\left(\frac{\tau^2 r^2}{4t}+1\right)(t+u)\right]\ ,
\label{140}
 \eeqn
 we arrive at the longitudinal cross section of
diffractive DY reaction,
 \beqn
\left.
\frac{d\sigma_{sd}^{DY(L)}}
{dM^2\,dx_1\,dp_T^2}\right|_{p_T=0}&=&
\frac{\sigma_0^2}{x_1}\,
\frac{2\alpha^2_{em}}{3(8\pi)^3}
\int\limits_{x_1}^1 d\alpha\,
\frac{(1-\alpha)^2}{\alpha}\,
F_2(x_1/\alpha)
\int\limits_0^\infty du\,e^{-u}
\nonumber\\ &\times&
\frac{1}{\tau^2}\,
\left[\frac{1+2w_1-4w_2}{1+4z} +
\frac{1+2w_3-4w_4}{(1+3z)(1+z)}
\right].
\label{150}
 \eeqn

The cross section of radiation of transversely polarized dileptons summed
over polarizations reads
 \beqn
&&
\left.
\frac{d\sigma_{sd}^{DY(T)}}
{dM^2\,dx_1\,dp_T^2}\right|_{p_T=0}=
\frac{\sigma_0^2}{M^2\,x_1}\,
\frac{\alpha^2_{em}}{3(8\pi)^3}
\int\limits_{x_1}^1 \frac{d\alpha}{\alpha}\,F_2(x_1/\alpha)
\int\limits_0^\infty du\,e^{-u}
\nonumber\\ &\times&
\left\{ \alpha^4 \frac{m_q^2}{\tau^2}\,
\left[\frac{1+2w_1-4w_2}{1+4z}
+
\frac{1+2w_3-4w_4}{(1+3z)(1+z)}
\right]\right.
\nonumber\\ &+& \left.
\frac{1-\alpha+\alpha^2}{u}
\left[\frac{2+o_1 - 2o_2}{1+4z}+
\frac{2+o_3-2o_4}{(1+3z)(1+z)}
\right]\right\}
\label{155}
 \eeqn

 Here $x_1\equiv x_\gamma$ and $x_2$ are the standard DY variables,
 \beq
x_1 = \frac{2qP_1}{s}\ ;
\hspace{2cm}
x_2 = \frac{2qP_2}{s}\ ,
 \label{157}
 \eeq
 where $q$ and $P_{1,2}$ are the 4-momenta of the radiated dilepton, and
projectile/target protons respectively.
Other notations in (\ref{155}) are,\\
 \begin{minipage}{5cm}
 \beqn
z &=& \frac{\la r_{ch}^2\ra}{R_0^2}\ ;
\nonumber\\
w_i &=& \frac{1-v_i^2}{4v_i}\,
\ln\left(\frac{1+v_i}{1-v_i}\right)\ ;
\nonumber\\
\lambda_1^2&=&\frac{8}{R_0^2(1+4z)}\ ;
\nonumber\\
\lambda_3^2 &=& \frac{8}{R_0^2(1+3z)}\ ;
\nonumber
 \eeqn
\end{minipage}
\hspace*{0.5cm}
\begin{minipage}{8cm}
 \beqn
o_i &=& (1-v_i^2)(2w_i+1)\ ;
\nonumber\\
v_i &=& \sqrt{\frac{u\,\lambda_i^2/\tau^2 }
{4+u\,\lambda_i^2/\tau^2}}\ ;
\nonumber\\
\lambda_2^2 &=& \frac{4(1+2z)}{R_0^2(1+4z)}\ ;
\label{160}\\
\lambda_4^2 &=& \frac{4(1+2z)}{R_0^2(1+3z)(1+z)}\ ;
\nonumber
 \eeqn
 \end{minipage}\\[0.5cm]
 and $i=1,2,3,4$.

Since the forward cross section is known, the total diffractive
cross section can be estimated as,
 \beq
\frac{d\sigma^{DY}_{sd}}{dx_1}=
\frac{1}{B^{DY}_{sd}(s)}\,\left.
\frac{d\sigma^{DY}_{sd}}{dx_1\,dp_T^2}\right|_{p_T=0}\ ,
\label{170}
 \eeq
 where $B^{DY}_{sd}(s)$ is the slope of the $t$-dependence of the cross
section, which is similar to the corresponding slope measured in
diffractive DIS.

\section{Unitarity corrections}

While the radiated leptons do not interact with the target, the
accompanying quarks do, and they can easily initiate an inelastic
interaction followed by multiparticle production filling up the rapidity
gap. Therefore, the survival probability of the gap causes a suppression
for the diffractive cross section. This is easy to take into account in
the impact parameter representation. The amplitude Eq.~(\ref{30}) acquires
a factor,
 \beq
A_{if}(b)\Rightarrow A_{if}(b)\,
\left[1-{\rm Im} f^{pp}_{el}(b)\right]\ ,
 \label{172}
 \eeq
 where $f^{pp}_{el}(b)$ is the partial elastic amplitude.
After squaring this amplitude and integration over impact parameter
we arrive at the cross section which is different from one in
Eq.~(\ref{170}) by a suppression factor $K$, $d\sigma^{DY}_{sd}/dx_1
\Rightarrow K\,d\sigma^{DY}_{sd}/dx_1$, where \cite{kps}
 \beqn
K=\left\{1-{1\over{\pi}}\,\frac{\sigma^{pp}_{tot}(s)}
{B^{DY}_{sd}(s)+2B^{pp}_{el}(s)} + \frac{1}{(4\pi)^2}\,
\frac{\left[\sigma^{pp}_{tot}(s)\right]^2}
{B^{pp}_{el}(s) \left[B^{DY}_{sd}(s)+B^{pp}_{el}(s)
\right]}\right\}\ .
 \label{174}
 \eeqn
 Here the elastic slope depends on energy as
$B^{pp}_{el}(s)=B^0_{el}+2\,\alpha^\prime_{\Pom}\, \ln(s/s_0)$ with
$B^{0}_{el}=7.5\GeV^{-2}$, $s_0=1\GeV^2$. The slope of single-diffractive
DY cross section can be estimated as, $B^{DY}_{sd}(s)\approx \la
r_{ch}^2\ra/3 +2\,\alpha^\prime_{\Pom}\,\ln(s/s_0)$, where the proton mean
charge radius squared $\la r_{ch}^2\ra=0.8\fm^2$.

Notice that this eikonal expression for the gap survival probability is a
conservative estimate. Inclusion of corrections related to intermediate
diffractive excitations of the beam particle (so called Gribov corrections
\cite{gribov}) make the medium more transparent, i.e. increase the
survival probability of the rapidity gap \cite{kps}.

Our estimate Eq.~(\ref{174}) can be compared with the results of more
elaborate models \cite{glm,glmnp,kmr,kkmr} incorporating a part of the
Gribov corrections. The predicted suppression factors have similar orders
of magnitude. Notice that one can easily replace the suppression factor
(\ref{174}) by a preferable one.

\section{Diffractive vs Inclusive}

We should compare the diffractive DY cross section with the inclusive
one, which also can be calculated within the dipole approach
\cite{hir,kst1,krt3},
 \beqn
\frac{d\sigma_{inc}^{DY(L,T)}}
{dx_1\,dM^2} =
\frac{1}{M^2\,x_1}\,\frac{\alpha_{em}^2}{3\pi}
\int\limits_{x_1}^1
\frac{d\alpha}{\alpha}\,F_2(x_1/\alpha)
\int d^2r\,\left|\tilde\Phi^{L,T}(\vec r,\alpha)\right|^2\,
\sigma\left(\alpha r,x_2\right)
\label{175}
 \eeqn
 This simple formula reproduces with high precision the cross section
measured by E772 \cite{e772} and E866 \cite{e866} experiments at
Fermilab, and agrees quite well with the results obtained from
much more complicated NLO parton model calculations \cite{rpn}.

 Applying Eqs.~(\ref{110})-(\ref{140}) to Eq.~(\ref{175}) we get,
 \beqn
\frac{d\sigma_{inc}^{DY(L)}}
{dx_1\,dM^2} =
\frac{\sigma_0}{x_1}\,
\frac{\alpha^2_{em}}{24\pi^2}
\int\limits_{x_1}^1 d\alpha\,
\frac{(1-\alpha)^2}{\alpha}\,F_2(x_1/\alpha)
\int\limits_0^\infty du\,e^{-u}\,
\frac{1-2w}{\tau^2}\ ;
\label{180}
 \eeqn

 \beqn
\frac{d\sigma_{inc}^{DY(T)}}
{dx_1\,dM^2} &=&
\frac{\sigma_0}{x_1\,M^2}\,
\frac{\alpha^2_{em}}{48\pi^2}
\int\limits_{x_1}^1
\frac{d\alpha}{\alpha}\,F_2(x_1/\alpha)
\int\limits_0^\infty du\,e^{-u}\,
\nonumber\\ &\times&
\left[\frac{\alpha^4\,m_q^2}{\tau^2}\,(1-2w)
+ \frac{1-\alpha+\alpha^2}{u}\,(2-o)\right]\ .
\label{190}
 \eeqn
 The notations here are similar to (\ref{160}),\\
 \begin{minipage}{5cm}
 \beqn
o &=& (1-v^2)(2w+1)\ ;
\nonumber\\
v &=& \sqrt{\frac{u\,\lambda^2/\tau^2 }
{4+u\,\lambda^2/\tau^2}}\ ;
\nonumber
 \eeqn
\end{minipage}
\hspace*{0cm}
\begin{minipage}{8.5cm}
 \beqn
w &=& \frac{1-v^2}{4v}\,
\ln\left(\frac{1+v}{1-v}\right)\ ;
\nonumber\\
\lambda^2&=&\frac{4}{R_0^2}\ ;
 \label{200}
 \eeqn
 \end{minipage}

We perform our calculations with the proton structure function
parametrized as,
 \beq
F_2(x,Q^2) = A(x)\left[\frac{\ln(Q^2/\Lambda^2)}
{\ln(Q_0^2/\Lambda^2)}\right]^{B(x)}
\left[1+\frac{C(x)}{Q^2}\right]\ ,
\label{210}
 \eeq
 with $Q_0^2=20\GeV^2$ and $\Lambda=0.25\GeV$. This parametrization was
fitted to available data for $F_2(x,Q^2)$ from different
experiments. The form and fitted parameters for the functions
$A(x),\ B(x)$ and $C(x)$ can be found in the Appendix of
\cite{smc}.

For the parameters of the dipole cross section, Eq.~(\ref{110}),
we use the results of the fit of Ref. \cite{gbw} to HERA data:
$\sigma_0=23.03\fm$; $R_0(x_2) = 0.4\fm\times(x_2/x_0)^{0.144}$,
where $x_0=3.04\times10^{-4}$.

\begin{figure}[h]
 \hspace*{3cm}
 \includegraphics[width=8cm]{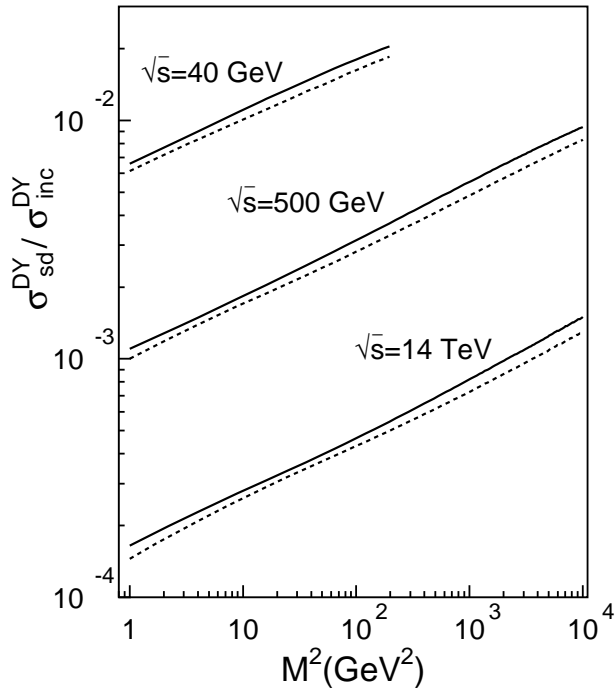}
\caption{Ratio of diffractive to inclusive DY cross sections
as function of $M^2$ at fixed $x_1=0.5$ (solid curves) and $x_1=0.9$
(dashed) and energies (from top to bottom) $\sqrt{s}=40\GeV,\
500\GeV$ and $14\TeV$.}
 \label{ratio}
 \end{figure}

We calculated the ratio of the diffractive to inclusive cross sections.
The results are plotted in Fig.~\ref{ratio} as function of the dilepton
effective mass squared at fixed $x_1=0.5,\ 0.9$ and energies
$\sqrt{s}=40\GeV,\ 500\GeV$ and $14\TeV$.
 In these calculations we summed the contributions of longitudinal and
transverse parts both in the diffractive and inclusive cross
sections.

The results depicted in Fig.~\ref{ratio} demonstrate unusual
effects. The diffractive-to-inclusive cross section ratio is
steeply falling with energy. This is counterintuitive, since
diffraction, which is proportional to the dipole cross section
squared, should rise with energy steeper than the total inclusive
cross section. At the same time, the ratio rises with the hard
scale, $M^2$. This also looks strange, since diffraction is
usually associated with soft interactions \cite{kp}.

To understand these remarkable features of DY diffraction we
should return back to the original diffractive amplitude,
Eq.~(\ref{40})-(\ref{50}). As we emphasized, the diffractive
amplitude is given by the difference between the cross sections of
the relevant Fock states with and without the dilepton. This is
explicitly incorporated into Eq.~(\ref{50}). Assuming $r\sim 1/M
\ll R_0(x_2)$ we can expand this cross section difference as,
 \beq
\sigma(\vec R)-\sigma(\vec R-\alpha\vec r)=
\frac{2\alpha\,\sigma_0}{R_0^2(x_2)}\,
e^{-R^2/R_0^2(x_2)}\,\left(\vec r\cdot\vec R\right)
+ O(r^2)\ .
\label{220}
 \eeq
 This is an interesting result: the amplitude is linear in $r$, i.e.
the diffractive cross section is quadratic function of $r$. This is
different from diffraction in DIS where the cross section is $\propto
r^4$. The latter is predominantly a soft process, since the end-point
$\bar qq$ fluctuations, $\alpha\to 0,\ 1$, have no scale dependence,
only their weight is $\propto 1/Q^2$. Therefore these soft
fluctuations dominate the diffractive DIS cross section
\cite{kp,diff}. However, since diffractive DY cross section is
$\propto r^2$, soft and hard interactions contribute on the same
footing, and their interplay does not depend on the scale, similar to
inclusive DY or DIS.

 Moreover, expression (\ref{240}) leads in Eq.~(\ref{80}) to the same
integral over $r$ and $\alpha$ as in the inclusive cross section
Eq.~(\ref{175}). Therefore, all energy and scale dependence of the
diffractive-to-inclusive cross section ratio comes via the $x_2$-dependent
factor,
 \beq
\frac{\sigma^{DY}_{sd}}{\sigma^{DY}_{inc}}\ \propto\
\frac{1}{R_0^2(x_2)}\, e^{-2R^2/R_0^2(x_2)}\ .
\label{240}
 \eeq
 Since for light hadrons $R_0^2(x_2)<2R^2$, this expression rises with
$R_0$, i.e. with $x_2=M^2/x_1s$. This explains why the ratio depicted in
Fig.~\ref{ratio} falls down with energy, but rises with $M^2$. Note that
the falling energy dependence is partially due to the absorptive
corrections (\ref{174}).

\begin{figure}[bht]
 \hspace*{3cm}
 \includegraphics[width=8cm]{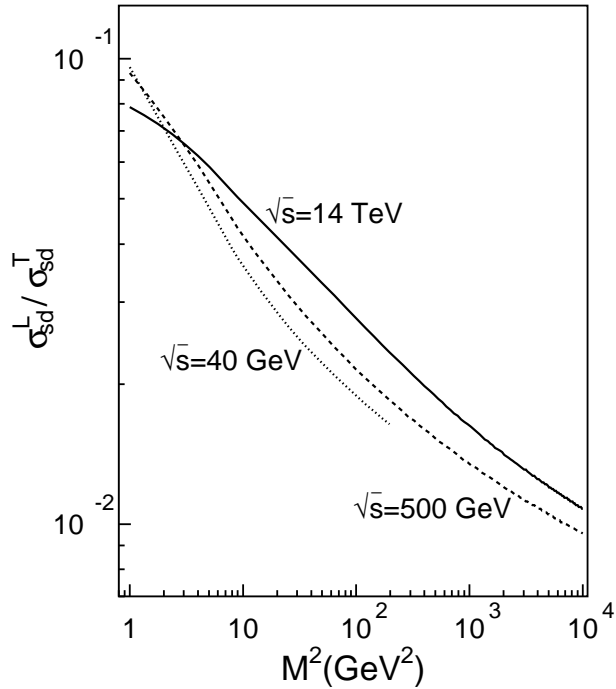}
\caption{Ratio of longitudinal to transverse parts of
diffractive DY cross
sections as function of $M^2$ at fixed $x_1=0.5$. The predictions are made
for energies, $\sqrt{s}=40\GeV$ (dotted), $\sqrt{s}=500\GeV$ (dashed),
and $\sqrt{s}=14\TeV$ (solid).}
 \label{L/T}
 \end{figure}

We found that both the diffractive and inclusive DY cross sections are
dominated by radiation of transversely polarized DY pairs. This is
demonstrated in Fig.~\ref{L/T} where we plotted the ratios of the
longitudinal to transverse cross sections for inclusive diffractive
DY reaction as function of $M^2$ at different energies.
 These ratios turn out to be very similar to that for inclusive DY 
processes.

 We see that the diffractive ratios are close to the inclusive ones. They
do not exhibit any strong dependence on energy, but fall down at
harder scales, similar to DIS \cite{krt2}. This fall is related to
the suppression of the end-point, $\alpha\to1$, of longitudinally
polarized DY fluctuations in equation(\ref{120}).

\section{Summary}

Contrary to the simple intuition based on QCD factorization,
diffractive DY processes have properties quite different from what
is known about DIS and nonabelian radiation.
\begin{itemize}

\item A quark cannot radiate noninteracting particles (photons, dileptons,
gauge bosons, Higgs, etc.) diffractively in the forward direction, i.e.
without gaining any momentum transfer. This is a manifestation of the
general principle: if a charge is accelerated and then immediately
decelerated, these two sources of radiation interfere destructively and
cancel each other \cite{kz,feri}. This can also be interpreted in terms of
the Landau-Pomeranchuk principle: radiation at times longer than the time
interval of the interaction depends only on the strength of the whole
kick, but does not resolve its structure (a single or multiple kicks).

The nonabelian case, QCD, is different: a quark can radiate gluons
diffractively in the forward direction\footnote{This is a higher order
effect. In Born approximation for inelastic collisions the forward
gluon radiation vanishes \cite{gb}.}. This happens due to possibility
of interaction between the radiated gluon and the target.

\item Nevertheless, a hadron can radiate abelian fields diffractively
and in the forward direction. This process involves the spectator
quarks as is illustrated in Fig.~\ref{graphs3}, and for this reason
the cross section depends on the hadronic size and rises with it. This
is an apparent and strong breakdown of factorization.

The situation in the nonabelian case is different. The main
contribution to the cross section comes from diffractive gluon
radiation by a single quark. Interaction with the spectator quarks
has little impact and vanishes at large separations.

\item The ratio of the cross section of diffractive radiation of abelian
fields to the inclusive one is steeply falling with energy as
depicted in Fig.~\ref{ratio}. This occurs even without unitarity
corrections which make the fall even steeper, and is a result of
the saturated shape of the dipole cross section. If this cross
section was rising like $\sigma(r)\propto r^2$, the ratio would be
energy independent (up to unitarity corrections). This property
is strikingly opposite to what would follow from factorization 
\cite{landshoff}.

This again is different in the nonabelian case, since here without
unitarity corrections the diffractive cross section rises faster
than the inclusive one. This happens because the diffraction cross
section is a quadratic function of the dipole cross section, while
the inclusive is linear .

This is also different from DIS where the diffraction to inclusive ratio
rises with energy.

\item Unexpectedly, the fraction of diffractive events to the total
inclusive cross section of abelian radiation rises with the dilepton
effective mass $M$ (see Fig.~\ref{ratio}).  This effect has the same
origin as the energy dependence, the dipole cross section which levels off
at large separations.  On the contrary, the fraction of diffraction in the
total DIS cross section is slowly, logarithmically, falling with the hard
scale. Same would be true for Drell-Yan diffraction if factorization were 
correct.

The rise of diffraction with the hard scale looks
counterintuitive, since diffraction is usually associated with
soft interactions. However, that is not true for diffractive DY
processes.

\item Hard and soft interactions contribute to DY diffraction on the same
footing, and their ratio is scale independent like in inclusive DY
process. This is a result of the specific property of DY
diffraction: its cross section is a linear, rather than quadratic
function of the dipole cross section.

On the contrary, diffractive DIS is predominantly a soft process, because
its cross section is proportional to $\sigma^2(r)$

\item The main features of our results for Drell-Yan reaction are valid for 
other diffractive abelian processes, like production of direct photons, 
Higgsstrahlung, radiation of $Z$ and $W$ bosons. One should not rely on QCD 
factorization for these reactions.

\item Measurement of diffractive DY processes in a wide energy range from
fixed target experiments up to the modern colliders, RHIC, Tevatron and
LHC, would provide a precious direct probe for the behavior of the dipole
cross section at large distances, above the saturated radius, $r>R_0$.

\end{itemize}

\bigskip

{\bf Acknowledgments:} We are grateful to Hans-J\"urgen Pirner for useful
discussions. A.V.T. thanks the Physics Department of the University
Federico Santa Maria for the hospitality. This work is supported in part
by Fondecyt (Chile) grants, numbers 1030355, 1050519, 1050589 and 7050175,
and by DFG (Germany)  grant PI182/3-1.

\end{document}